\documentclass{amsart}

\usepackage{amsmath}
\usepackage{graphicx}

\newcommand{\boundary}{\mathrm{boundary}}
\newcommand{\pop}{\mathrm{pop}}
\renewcommand{\deg}{\mathrm{deg}}
\newcommand{\area}{\mathrm{area}}
\newcommand{\dem}{\mathrm{dem}}
\newcommand{\repub}{\mathrm{repub}}

\newcommand{\vd}{\mathrm{votes}_D}
\newcommand{\vr}{\mathrm{votes}_{R}}
\renewcommand{\d}{\partial}
\newcommand{\J}{J}
\newcommand{\JJ}{J}
\newcommand{\dist}{\mathcal{R}}
\newcommand{\distGood}{\mathcal{R}_{connected}}
\newcommand{\pr}{\mathcal{P}}
\renewcommand{\Pr}{\pr_{\lambda,\beta}}

\title[Redistricting and the Will of the People]{Redistricting and the Will of the People\\{\tiny Preliminary
    Version - October 29th, 2014}}
\author{Jonathan C. Mattingly}
\address{Jonathan Mattingly, Departments of Mathematics and Statistical Science, Duke 
  University, Durham NC 27708}
\email{jonm@math.duke.edu}
\author{Christy Vaughn}
\address{Christy Vaughn, Department of Mathematics, Duke 
  University, Durham NC 27708}
\begin{document}
\maketitle

\begin{abstract}
  We introduce a non-partisan probability distribution on congressional
  redistricting of North Carolina which emphasizes the equal partition of the
  population and the compactness of districts. When random districts are drawn and the results of the 2012 election were re-tabulated under the drawn districtings, we find that an average of 7.6 democratic representatives
  are elected. 95\% of the randomly sampled redistrictings produced
  between 6 and 9 Democrats. Both of these facts are in stark contrast with the 4 Democrats
  elected in the 2012 elections with the same vote counts. This brings into serious
  question the idea that such elections represent the ``will of the
  people.'' It underlines the ability of redistricting to undermine
  the democratic process, while on the face allowing democracy to
  proceed.
\end{abstract}

\section{Introduction}

Democracy is typically equated with expressing the will of the people
through government.  Perceived failures of democracy in representative
governments are typically attributed to the voice of the people being
muted and obstructed by the actions of special interests or the sheer
size of government. The implication being that the voice and will of the
people exists as a clear well defined voice which only needs to be
better heard.

Yet the will of the people is not monolithic. It is not always so
simple to obtain a consensus or even a clear majority opinion. We rely on our
elections as a proxy to express our collective opinions and our political
will.  Of course, Arrow's Impossibility Theorem guarantees that all
electoral systems have paradoxes. However, there is a more profound
problem with this notion; the very concept of a clear majority voice is
problematic. 
In the United States, the federalist system, on the national
level, and district representation schemes, on the municipal level,
acknowledge that the people's voice is geographically diverse and
that we value the expression of that diversity in our government. 
It is reasonable to
ask how singular is the received ``will of the people'' when it is
filtered through geographically based districts. We take election
results to
give the elected
officials a mandate to act in the people's name. 
How sensitive are
the election results to the choice of districts? By extension, how
sensitive to the choice of redistricting is the received impression of
the ``will of the people?''
Our results show that  the ``will of the people'' is not a single election outcome but rather a
distribution of possible outcomes. The exact same vote counts can lead
to drastically different outcomes depending on the choice of
districts.

With the increased insertion of politics into the congressional redistricting
process, exploring these questions in the context of house congressional
districts seems particularly important and timely. The 2010
redistricting of North Carolina is a useful example and testing ground
for this general line of inquiry. Most would agree that politics had
a hand in the North Carolina 
redistricting process. The motivations were diverse. The twelfth 
district was drawn to create a majority black district. Others seemingly 
were drawn to split and pack different voting blocks to diminish their 
political power, particularly those of the democratic party. The
question remains of how large was the effect of the redistricting on
the outcome.


In the 2012 congressional elections, which were based on the 2010
districts, four out of the thirteen 
congressional seats were filled by Democrats. 
Yet in seeming contradiction, the majority of
votes were cast for Democratic candidates on the state wide level. The election results hinged on the
geographic positioning of congressional districts.
While this outcome is clearly the result
of politically drawn districts, perhaps it is not the result of
excessive tampering. Our country has a long
history of balancing the rights of urban areas with high population
with those of more rural, less populated areas. Our federalist and
electoral structures enshrined the idea that the
one-person-one-vote ideal could be modified to
support other objectives, particularly that of regionalism. It might be that in North Carolina the
subversion of the results of the global vote count would happen
in any redistricting which balances the representation of the urban with the
rural or the beach, with the mountains, and each with the Piedmont.  Maybe
the vast majority of reasonable districts which one might draw would
have these issues due to the geography of the population's
distribution. We are left asking the basic question: ``How much does
the outcome depend on the choice of districts?'' This can be further
refined by asking ``what are the outcomes for a typical choice of
districts?'' or ``When should a redistricting be considered
outside the norm?'' These last two refinements require some way
of quantifying what the typical outcomes are for a given set of
votes. This turns the usual election procedure on its head. We are
interested in fixing the votes and then changing the redistricting and
observing how the results change. Since we will explore these
questions in the context of the American political system, we will
assume that people vote for parties, not people, which is of course not
true. However, in these polarized times 
it is not the worst approximation. We still find the results 
extremely illuminating.

Once one accepts that the
expressed ``people's will'' is not a single outcome but a distribution
of outcomes depending on the redistricting, it expands the realm of possibility in evaluating
redistrictings. The principle goal of this article is to construct an
appropriate probability distribution on all redistrictings 
and explore its implications. Our probability distribution will be
nonpartisan in that it will only make use of the distribution of the
population and not any information about party affiliation,
historical voting patterns, race or socioeconomic class.

We do not intend for this work to be the definitive answer in this
direction. Rather, we have erred on the side of simplicity by
constructing a probability distribution which only considers the
compactness of districts and the degree to which the population is
partitioned equally. More complicated procedures might value minority
representation or traditional political boundaries more. We simply
wish to show the utility of a simple probability distribution placed
on the space of redistrictings to illuminate the degree to which the
outcome of an election depends on the choice of districts and if a
given redistricting produces representative results. Using such
a distribution, one begins to obtain a feel for the degree to which
the perceived will of the people fluctuates and how typical, and
perhaps ``fair'', a given redistricting is.

In Section~\ref{sec:resultsSum} we describe our main
results. In Section~\ref{sec:ConstructingMeasure}, we detail the
construction of a probability measure on possible redestrictings. In
Section~\ref{sec:motivating-examples}, we explain a number of
illustrative examples designed for those not familiar with
the types of Gibbs probability measures constructed in
Section~\ref{sec:ConstructingMeasure}. In Section~\ref{Sampling}, we
discuss the algorithm used to draw example redistrictings from our
probability measure. In Section~\ref{sec:calibrating}, we discuss how
the model is calibrated to produce acceptable results. In
Section~\ref{sec:racial}, we make a small digression to discuss the
racial make up of districts produced by our measure.  Finally, in
Section~\ref{sec:conclusions}, we make some concluding remarks and
discuss some future directions.

\section{Summary of Results}
\label{sec:resultsSum}
\begin{figure}[ht]\label{fig:firstHist}
  \centering
\includegraphics{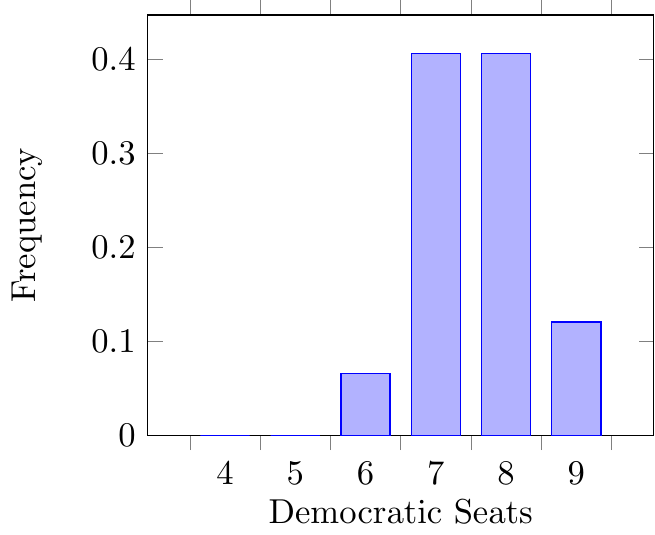}
\caption{Histogram of Democratic seats counts  after over 300
  samples. ($\lambda=0.3$ and Long period. See Section~\ref{sec:calibrating}) }
 \end{figure}

To examine the effect of the choices of districts on North Carolina
Congressional elections, we will place a probability distribution on
the space of all possible redistricting plans of the state into
thirteen federal Congressional districts.  We will then choose a
redistricting according to this probability and then rerun the North
Carolina congressional elections from 2012 using the actual votes
recorded in each voting tabulation district (VTD). This will produce a
winner from each of the thirteen congressional districts. We will
record the number of democrats (or equivalently republicans) elected
and then repeat the procedure drawing a new, fresh redistricting from
our probability distribution and again recording the outcome of the
election with these new districts. After many such draws, we obtain a
histogram showing the distribution of outcomes. It shows the
sensitivity of the results to the districts chosen. The results of
this procedure for our principle model are given by the histogram
given in Figure~\ref{fig:firstHist} which shows fraction of different
numbers of Democratic representatives obtained by sampling about 100
random redistrictings from our random distribution. This distribution
 gives a quantitative measure of the ``will of the people'' in a given
election.  Those in search of a single number might well take the mean
(7.6 Democratic representatives) or the median (7 Democratic
representatives). However, the entirety of the distribution gives
more information.  Over 50\% of the samples produce either 7 or 8
Democratic representative. All of the samples produce between 6 and 9
Democratic representatives. These results should be compared with the current
North Carolina house delegation which has only 4 Democratic
representatives.

In light of these results, it might be
reasonable to accept redistrictings which produce outcomes within one standard deviation of the mean to be
truly representative while those outside to be suspect and
not representative of the will of the people. Not once in our run were 4 or less
Democratic seats produced. While it is possible, our results show that
it is extremely unlikely that a random redistricting, chosen according
to our nonpartisan probability distribution, would produce 5 or less
Democratic seats if the actual vote counts from the 2012 election are used.

 In building our probability measure on redistrictings, we will only
 include the three main legal requirements of a redistricting
 plan. First, the districts should be connected. Second, they should
 come as close as possible to having equal number of people. Lastly,
 they should be as compact as possible. Our procedure will tacitly
 include the fact that redistricting should have some relationship to
 historical districts and communities; however, we postpone that
 discussion to later. (See the end of Section~\ref{Sampling}.) We
 emphasis that no information about party affiliation or the vote
 counts from any election were used in generating the probability
 distribution. Furthermore, the results in Figure~\ref{fig:firstHist}
 used the actual vote tally from the 2012 congressional election.

Given the suppressing nature of the results, it is natural to ask if
the districts produced are
reasonable. Figures~\ref{fig:sampeRedistricting-1}--\ref{fig:sampeRedistricting-3}
give example redistrictings chosen randomly from the collection we
produced.  The law mandates that each district should have as close to
$\frac{1}{13}$ of the total population as possible. All of the districts
considered in making Figure~\ref{fig:firstHist} had a relative deviation
of less than $0.7\%$ from the ideal of $\frac{1}{13}$. $50\%$ percent had
a relative deviation less than $0.09\%$. This compares favorably with
the districting currently under use which has relative deviations less than $0.7\%$.

Our goal is neither to provide a method for producing usable
redistricting nor a  definitive model for the likelihood
of a given redistricting, but rather to provide a simple model which
can be understood and which can shed light on the ``the people's will.''
Nonetheless, our model did contain a tunable parameter, denoted $\lambda$,
which varies between zero and one
and measures the relative weight given to the constraints of equal
division of population versus the compactness of districts. When
$\lambda$ is close to zero, all of the weight is given to the
compactness, while when $\lambda$ is close to one, all of the weight is
given to the division of population. Since the two effects
are not necessarily on the same scale, $\lambda$ equal one-half does
not necessarily represent the equal balancing of the effects. One must
look at the sample redistrictings to calibrate the parameters. 
The data quoted above and shown in Figure~\ref{fig:firstHist} used
$\lambda=0.3$ which was chosen because the districts produced with this value were overwhelmingly
better than the currently used districts at splitting the population evenly between the thirteen
districts and at being more compact. 

\begin{figure}[ht]\label{fig:summaryAll}
   \centering 
\includegraphics{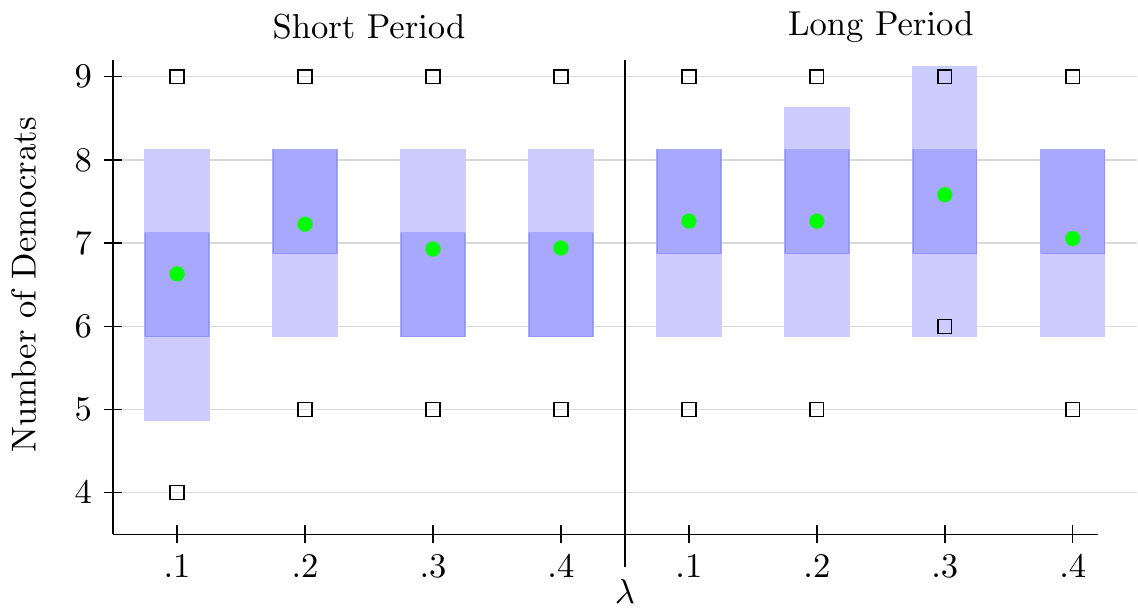}
\caption{Plotted are the summary of the elections for four different 
  values of the parameter $\lambda$ used in defining our probability 
  distribution of redistrictings.  Each of the four different values 
  of $\lambda$ is under a shorter and longer heating/cooling cycles.  The solid green dot is the 
  median. The dark blue box is centered on the median to contain 50\%
  of the points. The lighter blue box is centered to contain 90\% of 
  the data. The hollow squares give the max and min values. } 
\end{figure}

We
further discuss selecting a value of $\lambda$ in
Section~\ref{sec:calibrating}. Nonetheless, in Figure~\ref{fig:summaryAll} we summarize the results with four
different values of $\lambda$ as well as two different versions of the
method used to sample the probability distribution: ``Long
Period'' and ``Short
Period'' methods.
The ``Long
Period'' method is preferred and was used in
Figure~\ref{fig:firstHist}  and the data quoted above. The ``Short
Period'' method allows for more data to be collected but is less
effective at drawing from the desired probability distribution. A discussion
of the issues involved is given in Section~\ref{Sampling} and Section~\ref{sec:calibrating}.

What is striking about Figure~\ref{fig:summaryAll} is that the basic
conclusions are relatively
insensitive to the choice of $\lambda$ or the details of how we sample
the system. All of the distributions we produced have a mean
around seven. All of the distributions are fairly concentrated
around the mean with 50\% of the values concentrated on only two
values, one always being seven. It is also worth noting that all but one of
the distributions never produced only four democratic representatives
which was the number of democratic representatives elected in the 2012
elections. The one that did, did so in less than 5\% of the samples
and was the distribution possessing fewest desirable properties of all
those considered. 
Hence a randomly chosen redistricting from any
of the probability distribution considered is all but
guaranteed to produce drastically different results from the 2012
elections while using the
exact same votes as the 2012 election. 

It is clear that the current situation of only four democratic
representatives is not representative of the will of the people. The
results of the election seem to have very little to do with the will
of the people.
It is worth recalling that our probability distribution on all
possible redistrictings was agnostic relative to the different
political parties. It uses no information about the number of
registered democrats or republicans in a district nor its racial or
socioeconomic make up. We are simply drawing, in an unbiased way,
redistrictings, favoring those which are compact  and those which have
approximately 1/13th of the state's population. The result left the
authors wondering, ``Is this democracy?''


%
%

\section{A probability distribution on reasonable redistricting}
\label{sec:ConstructingMeasure}
We will represent the state of North Carolina as a graph $G$ with
edges $E$ and vertices $V$. Each vertex represents a Voting
Tabulation District (VTD) and an edge between two vertices exists if
the two VTDs are adjacent on the map. This graph representing the
North Carolina voting landscape has over 2500 vertices and over
8000 edges.\footnote{See Section~\ref{sec:Technical Notes} for some technical notes about the construction of this graph.}


A redistricting plan is a function from the set of vertices to the
integers from one to thirteen (since North Carolina has thirteen seats
in the house of representatives). More formally, recalling that $V$ was
the set of vertices, we will represent a redistricting plan by a
function $\xi: V \rightarrow \{1,2,\dots,13\}$.
We let $\dist$ denote the space of all redistricting plans. If $\xi(v) =i$ for some $v \in V$
then the VTD represented by vertex $v$ is in district $i$.
Similarly for $i \in \{1,2,\dots, 13\}$ and a
plan $\xi$, the
$i$ district which we will denote by $D_i(\xi)$ is given by $\{v \in
V: \xi(v)=i\}$.  We wish to consider  redistricting plans $\xi$ such
that each district $D_i(\xi)$ is a single connected component. We
will denote the collection of such redistricting plans by $\distGood
\subset \dist$.

As already described in Section~\ref{sec:resultsSum}, our goal is to 
produce a probability distribution on the space of 
redistrictings. To define our probability measure on the space of redistrictings, we
will first assign a score to each redistricting plan $\xi$ according to how well it
satisfies our ideals, with lower scores preferred. 
To describe the score function on the space of redistricting
plans, we need to attach to our graph $G=(V,E)$ some data which gives
relevant features of each VTD.
We define the positive functions
$\pop(v)$ and $\area(v)$ for a vertex $v \in V$ as respectively the
population and geographic area of the VTD
associated with the vertex $v$. We extend these
functions to a collection of vertices $A \subset V$ by
\begin{align}\label{eq:PopArea}
  \pop(A) = \sum_{v \in A} \pop(v) \qquad\text{and}\qquad \area(A) =
  \sum_{v \in A} \area(v)\,.
\end{align}

We will think of the boundary of a district
$D_i(\xi)$ to be the subset of the edges $E$ which connect vertices
inside of $D_i(\xi)$ to vertices outside of $D_i(\xi)$. We will write
$\d D_i(\xi)$ for the boundary of the district $D_i(\xi)$. Since we
want to include the exterior boundary of each district (the section
bordering an adjacent state or the ocean), we add to $V$ the vertex $o$
which represents the ``outside'' and connect it with an edge to each
vertex representing a VTD which is on the boundary of the state. We
will always assume that any redistricting $\xi$ always satisfies
$\xi(v)=0$ if and only $v=o$. 
Since $\xi$ always satisfies $\xi(o)=0$ and hence $o \not \in
D_i(\xi)$ for $i \geq 1$, it does not matter that we have not defined
$\area(o)$ or $\pop(o)$ as $o$ is never included in the districts.

Given an edge $e \in E$
which connects the two vertices $v, \tilde v \in V$, we define
$\boundary(e)$ to be the length of common border of the VTDs
associated with the vertex $v$ and $\tilde v$.  As before, we extend
the definition to the boundary of a set of edges
$B\subset E$ by 
\begin{align}
\label{eq:Boundary}
    \boundary(B) = \sum_{e \in B} \boundary(e)\,.
\end{align}

With these preliminaries out of the way, we return to defining the
score functions used to assess the goodness of a redistricting.  We will construct our
total score function as a convex combination of two terms: a
population score $\JJ_{pop}$ and a compactness score
$\JJ_{compact}$.
We define the population score by
\begin{align*}
   \JJ_{pop}(\xi) = c_{pop}\sum_{i=1}^{13} \Big(\pop(D_i(\xi)) - \frac{N_{pop}}{13}\Big)^2
\end{align*}
where $N_{pop}$ is the total population of North Carolina,
$\pop(D_i(\xi))$ is the population of the district $D_i(\xi)$ as
defined in \eqref{eq:PopArea}, and $c_{pop}$ is a positive constant
which is used to make the size of the two score terms comparable.

We define the compactness score $\JJ_{compact}$ as a ratio of the sum of the
perimeter to the total area of each district. This ratio, often
referred to as the ``isoperimetric constant'' of a region, is
minimized for a circle which is the most compact shape. Hence we define
\begin{align*}
  \JJ_{compact}(\xi)= c_{compact} \sum_{i=1}^{13}\frac{\big[\boundary(\d D_i(\xi))\big]^2}{\area(D_i(\xi))}\,.
\end{align*}
where $\d D_i(\xi)$ is the set of edges which define the boundary,
$\boundary(\d D_i(\xi))$ is the length of the boundary of district
$D_i$ and $\area( D_i(\xi))$ is its area. As before $c_{compact}$ is
used to make the size of $ \JJ_{compact}$ and $\JJ_{pop}$
comparable. After some experimentation, we found that $c_{pop}=1/5000$ and $c_{compact}=2000$ bring $ \JJ_{compact}$ and $\JJ_{pop}$ to about the same scale. This compactness measure is one of two measures often used in the legal
literature where it is referred to as \textit{the parameter
  score}\cite{Pildes_Niemi_1993,Practice_Hebert_2010}. A second
measure often used is the ratio of the area of the district to that of
the smallest circle which contains the district. This second measure,
usually referred to as \textit{the dispersion score}, is more
sensitive to overly elongated districts though the parameter score
also penalizes them. We will not use the dispersion score since the
spatial location of each VTD was not readily available.

Recall that $\distGood$ was the collection of
redistricting policies in which all districts are connected.
For any $\lambda \in [0,1]$ and $\xi \in \distGood$, we define
$\JJ_\lambda$ by
\begin{align}
  \label{eq:J}
  \JJ_\lambda(\xi) = \lambda \JJ_{pop}(\xi) + (1-\lambda) \JJ_{compact}(\xi) \,.
\end{align}
The parameter $\lambda$  dictates the balance between
the two energies. 
The lower $\JJ_\lambda(\xi)$ is,
the more evenly distributed the population is and the more compact the
districts $D_i(\xi)$ are.  

So far we have defined $\JJ_\lambda(\xi)$ for $\xi \in \distGood$. We
now extended the definition by
\begin{equation}
  \label{eq:1}
  \JJ_\lambda =
  \begin{cases}
     \lambda \JJ_{pop}(\xi) + (1-\lambda) \JJ_{compact}(\xi)  &
     \xi \in \distGood\\
     \infty &\xi \not \in \distGood
  \end{cases}
\end{equation}
We will see that if $\JJ_\lambda(\xi)=\infty$, then the probability
that the redistricting $\xi$ is considered will be zero.

Next for all $\beta >0$ and $\lambda \in [0,1]$, we define the probability measure $\mathcal{P}_{\lambda,\beta}$ on the space
of redistrictings $\dist$ by
\begin{align}
  \label{eq:Prob}
  \Pr(\xi) = \frac{e^{- \beta \JJ_\lambda(\xi)}}{\mathcal{Z}_{\lambda,\beta}}
\end{align}
where $\mathcal{Z}_{\lambda,\beta}$ is simply the normalization
constant defined so that $\Pr(\dist)=1$. In other words,
\begin{align*}
    \mathcal{Z}_{\lambda,\beta} = \sum_{\xi \in \dist} e^{- \beta J_\lambda(\xi)}\,.
\end{align*}
Notice that as promised  if $\xi \not \in \distGood$ then $e^{-\beta
  J_\lambda(\xi)} = e^{-\infty}= 0$ and we see that
$\Pr(\distGood)=1$. Hence, all of the probability is concentrated on
redistrictings which have simply connected districts. The positive
constant $\beta$ is often called the ``inverse temperature'' in
analogy with statistical mechanics and gas dynamics. When $\beta$ is very small (the
high temperature regime), different elements of $\distGood$ have close
to equal probability. As $\beta$ increases (the ``temperature
decreases''), the measure concentrates the probability on the
redistrictings $\xi \in \dist$ which minimize $J_\lambda(\xi)$. After
some experimentation we found that $\beta=0.01$ produced districts
with reasonably well balanced populations and compact footprints when
compared to the districts currently used.

\subsection{Estimating Boundary Sizes}

Unfortunately, not all of the data we need to define $\J_\lambda$ is
readily available. The information to define $\pop(v)$ and $\area(v)$
is publicly available for all of the VTDs in North Carolina \cite{NCPop} \cite{shapeFilesCongress}, while the
value for $\boundary(e)$ is not. In principle it could be obtained
from the map showing the North Carolina VTDs. However, we had no means
to efficiently automate this task. Since each of the VTDs is
relatively small and many make up a given congressional district, we
opted to employ a simple approximation. 

Since the area of a VTD and the number of neighboring VTDs are readily
available, we approximated the length of the shared boundary between
two VTDs as follows. By assuming each VTD is a circle, we expressed
the circumference as a function of the area by 
\begin{align*}
  \text{Circumference} = 2 \pi^{\frac32} \sqrt{\text{Area}} \,.
\end{align*}
Then if we further make the approximation that the circumference is
equally shared with each neighboring VTD, we can estimate the shared
boundary between two adjacent VTDs as
\begin{align*}
  \text{Shared Boundary}=  2 \pi^{\frac32} \frac{\sqrt{\text{Area}}}{\text{Number of Neighbors}}
\end{align*}
Since this approximation could be centered on either of the two
vertices which make up the edge, we average the two answers. Denoting
the degree of a vertex by $\deg(v)$ and noting that constants will not
change the relative size of the term we define
\begin{align}
  \label{eq:boudnaryAprox}
  \boundary(e)=\frac12\left( \frac{\sqrt{\area(v)}}{\deg(v)} +\frac{\sqrt{\area(v')}}{\deg(v')}  \right) 
\end{align}
if $e$ is an edge connecting vertices $v$ and $v'$ with neither edge
being the ``outside'' vertex $o$. If the edge $e$ connects an interior
vertex $v$ with the outside vertex $o$, we set
\begin{align}
  \label{eq:boudnaryAprox}
  \boundary(e)=\frac{\sqrt{\area(v)}}{\deg(v)}\
\end{align}
since the estimate centered at $o$ does not make sense.

\section{Motivating and Explanatory Examples}
\label{sec:motivating-examples}
To clarify concepts used to define a probability measure on the space of
redistrictings, we now give some elementary, illustrative
examples. Those familiar with the idea of a Gibbs measure can skip
ahead to the next section.

\begin{figure}[ht]
  \centering
\includegraphics{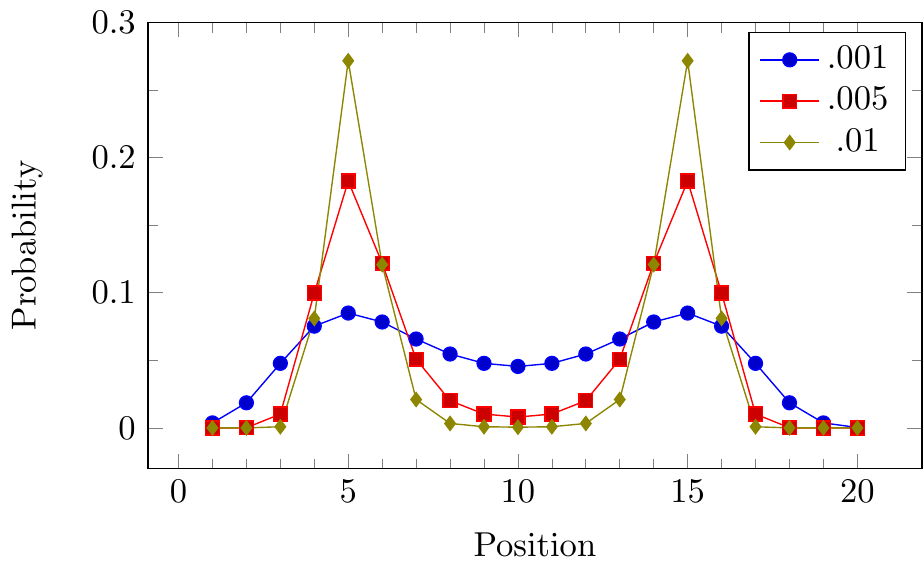}
\caption{Plot of the probability of the point $\{1,2,\dots,20\}$ for
  $\pr_\beta$ with $\beta$ equal to 0.001, 0.005, and 0.01.}
\label{fig:1D}
\end{figure}

To better understand the role of the inverse temperature $\beta$, we
begin by constructing a series of measures on the integers
$\{1,2,\dots,20\}$. As in the redistrictings setting, we begin by
defining a score function which we denote by $\mathcal{H}$. For any $i
\in \{1,2,\dots,20\}$ we define
\begin{align}\label{eq:H}
  \mathcal{H}(i) = (i-5)^2 \cdot  (i-15)^2 + 1
\end{align}
and the probability measure $\pr_{\beta}$ by
\begin{align*}
  \pr_{\beta}(x)=\frac{e^{-\beta \mathcal{H}(x)}}{Z_\beta}\qquad\text{where}\qquad  Z_\beta=\sum_{i=1}^{20} e^{-\beta \mathcal{H}(i)}\,.
\end{align*}
Figure~\ref{fig:1D} shows a plot of $\pr_\beta$ for $\beta$ equal to $.001$,
$.005$, and $.01$. Since $\mathcal{H}(x)$ is smallest when $x$ is 5 or
15, the measure tends to concentrate around these values. However, the
degree it does so is governed by the inverse temperature $\beta$. When
$\beta$ is very small, the probability is more evenly distributed. When
$\beta$ is larger, the probability is highly concentrated around the
minimum of $\mathcal{H}$.
%
%


\begin{figure}[ht]
  \centering
  %
\includegraphics{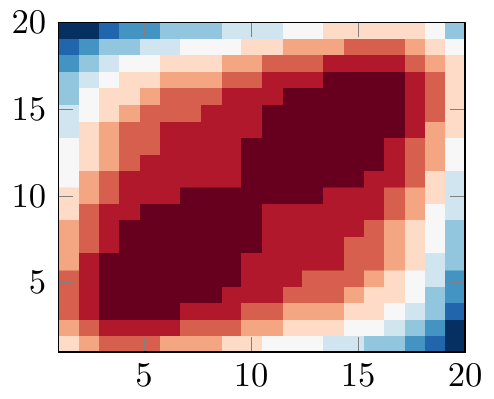}
\includegraphics{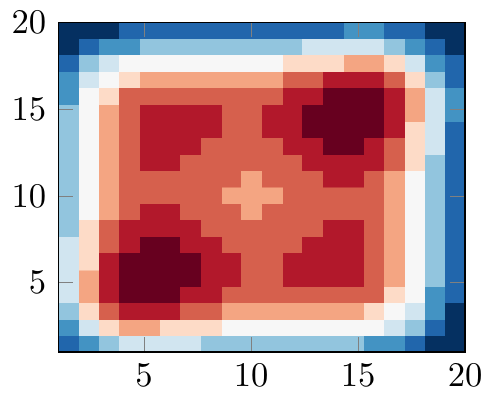}
\\
\includegraphics{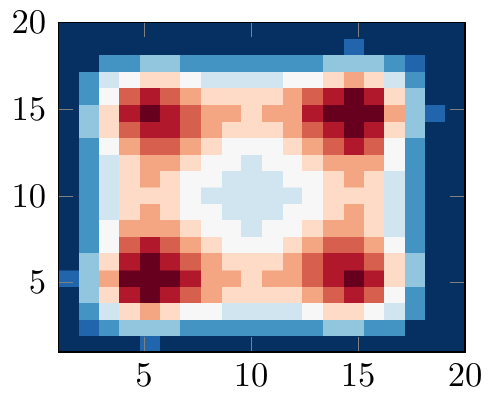}
%
\includegraphics{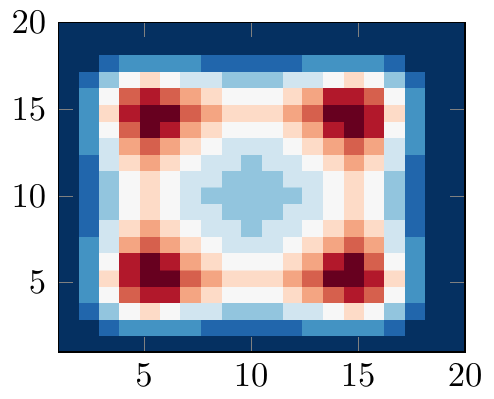}
   \caption{Heat maps of $\Pr(i,k)$ with red denoting relatively high 
    probability and blue relatively low. All plots use 
    $\beta=0.001$. The values of $\lambda$ are 0.95 (upper right),
    0.75 (lower right), 0.25 (lower left), and 0.05 (upper left)}
  \label{fig:2D}
\end{figure}

Now lets consider a second example to explore the roll of the relative
weighting parameter $\lambda$ in \eqref{eq:J}. Again we will construct
an artificial example in lower dimensions where the effects are easier
to visualize. We will place a probability measure $\Pr$ on
$\{1,\cdots,20\}^2$, the space of all
pairs $(i,k)$ where $i,k \in \{1,\dots,20\}$. As before we will define 
\begin{align*}
  \Pr(i,k)= \frac{e^{-\beta \mathcal{H}_\lambda(i,k)}}{Z_{\beta,\lambda}}
\end{align*}
where the score function $\mathcal{H}_\lambda$ is defined for $\lambda
\in [0,1]$ by
  \begin{align}
    \label{eq:Hlambda}
    \mathcal{H}_\lambda(i,k) = \lambda \mathcal{H}_{single}(i,k) + (1-\lambda)\mathcal{H}_{pair}
  \end{align}
with $\mathcal{H}_{single}(i,k)=\mathcal{H}(i)+\mathcal{H}(k)$ where
$\mathcal{H}$ was defined in \eqref{eq:H}, $\mathcal{H}_{pair}$ is
defined by $\mathcal{H}_{pair}(i,j)=(i-k)^2$, and as before the
normalizing constant is given by
\begin{equation*}
Z_{\beta,\lambda}=\sum_{i,k=1}^{20} e^{-\beta \mathcal{H}_\lambda(i,k)}\,.
\end{equation*}
Figure~\ref{fig:2D} shows the probability in a two-dimensional ``heat
map'' for different values of $\lambda$ and $\beta=.001$. Red points
denote relatively high probability, white points denote probability in the
middle of the range, and blue points denote relatively low probabilities. The value
of $\lambda$ starts high in the upper right plot and decreases
clockwise to a lowest value in the upper left corner.
%

Notice how as $\lambda$ shifts, the qualitative features of the measure
shift. When $\lambda$ is small, the value of the score is dominated by
$\mathcal{H}_{pair}(i,k)$ which favors pairs where $i$ and $k$ are
close. When  $\lambda$ is large, $\mathcal{H}_{single}(i,k)$ dominates
and the most probability is placed on pairs near the points $\{(5,5),
(5,15), (15,5) (15,15)\}$. For intermediate values of $\lambda$, a
balance is struck between the two goals.

\section{Rerunning elections with the 2012 votes}

To any redistricting $\xi \in \dist$ we will associate an election
outcome using the actual 2012 votes in each VTD. We let $\vd(v)$ and
$\vr(v)$ denote respectively the number of democratic and
republican votes in the VTD associated with the vertex $v$ in the 2012
congressional elections as obtained from the North
Carolina Board of Elections \cite{VTD2012}.  Then as before, for a set
of vertices $A \subset V$ we define
\begin{align*}
   \vd(A)= \sum_{ v \in A} \vd(v) \qquad\text{and}\qquad    \vr(A)= \sum_{ v \in A}  \vr(v)\,. 
\end{align*}
Then for any redistricting $\xi$ we define 
\begin{align*}
  \dem(\xi)&=\#\{ i :  \vd(D_i(\xi)) >  \vr(D_i(\xi))\}\\
 \repub(\xi)&=\#\{ i :  \vd(D_i(\xi)) <  \vr(D_i(\xi))\} 
\end{align*}
where $\#\,C$ is the number of elements in the set $C$. Hence
$\dem(\xi)$ and $\repub(\xi)$ give the number of democratic and
republican congressional seats associated to a redistricting $\xi$ if
we assume that each voter votes for the party they did in the 2012 election.
 
We are then primarily interested in the distribution of $\dem(\Xi)$
(or equivalently $\repub(\Xi)$) when $\Xi$ is chosen randomly
according to the probability distribution $\Pr$. This will give some
understanding of the make up of the congressional delegation for
typical redistricting. It is also useful to record mean of $\dem(\Xi)$ which is defined by
\begin{align*}
  \mu_{dem} = \sum_{\xi \in \dist} \dem(\xi)
  \Pr(\xi)\,.
\end{align*}

\section{Sampling the probability measure $\Pr$}
\label{Sampling}
There are more then $13^{2500}\approx 7.2 \times 10^{2784}$ different
redistrictings in $\dist$. Current estimates on the number of atoms in
the universe range from $10^{78}$ to $10^{82}$ and the number of
seconds since the big bang at the creation of the universe is
estimated to be $4.3\times 10^{17}$ seconds. While there are
significantly less redistricting in $\distGood$ (the set of simply
connected redistrictings), it is certainly not practical to enumerate
the redistrictings to find those with the lowest values of
$\J_\lambda$ and hence the largest probability.

The standard and very effective way to escape this curse of
dimensionality is to use a Markov chain Monte Carlo (MCMC) algorithm
to sample from the probability distribution $\Pr$. The basic idea is
to define a random walk on $\distGood$ which has $\Pr$ as its globally attracting
stationary measure. We do this using the standard Metropolis-Hastings
algorithm which we now briefly explain.

The Metropolis-Hastings algorithm is designed to use one Markov transition
kernel $Q$ (the proposal chain) to sample from another Markov transition
kernel which has a unique stationary distribution $\mu$ (the target distribution).
$Q(\xi,\xi')$ gives the  probability of moving from the
redistricting $\xi$ to the redistricting $\xi'$ in the proposal Markov
chain and is assumed to be readily computable. We wish to use $Q$ to
draw a sample distributed according to $\mu$. 
The algorithm proceeds as follows:
\begin{enumerate}
\item Choose some initial state $\xi \in \dist$.
\item Propose a new state $\xi '$ with transition probabilities given
  by $Q(\xi,\xi')$.
\item Accept the proposed state with probability $p=\min
  \big(1,\frac{\mu(\xi ') q(\xi ', \xi)}{\mu(\xi) Q(\xi, \xi ')}\big)$.
\item Repeat steps 2 and 3.
\end{enumerate}
After an initial burn-in period, the stationary distribution of this
Markov chain matches the stationary measure $\mu$. Thus, the states can be treated as samples from the desired distribution.

The stationary measure we would like to sample is $\Pr$. Our initial
state is the districting that was used for the 2012 US House of
Representatives election. We define the proposal chain used for
proposing new redistricting in the following way:
\begin{enumerate}
\item Uniformly pick a conflicted edge at random. An edge, $e=(u,v)$
  is a conflicted edge if $\xi (u) \neq \xi(v)$, $\xi(u) \neq 0$,
  $\xi(v) \neq 0$.
\item For chosen edge $e=(u,v)$, with probability $\frac12$, either:
\begin{equation*}
   \xi '(w) = 
     \begin{cases}
       \xi(w) & w \neq u\\
       \xi(v)& u
     \end{cases}
\qquad\text{or}\qquad
\xi '(w) =
\begin{cases}
   \xi(w) & w \neq v\\
       \xi(u)& v 
\end{cases}
\end{equation*} 
\end{enumerate}
Let $con(\xi )$ be the number of conflicted edges for districting
$\xi$.  Then we have $q(\xi,\xi ')=\frac{1}{2 con(\xi)}$. The
acceptance probability is given by:

$$p=\min \big(1, \frac{con(\xi)}{con(\xi ')} e^{-\beta (\J_\lambda(\xi ')-\J_\lambda(\xi))} \big)$$
Recall that if a districting $\xi '$ is not connected, then
$\J_\lambda(\xi ')=\infty$. Thus, proposed redistrictings that are not
connected are never accepted.

After a burn in period, every $m$-th districting can be taken as a
sample from $\Pr$ for some $m$. If $m$ is long enough, the samples will
be essentially independent. In our test, we used $m=40,000$ and
$m=100,000$. The principle results quoted in
Section~\ref{sec:resultsSum} used the larger value. The fact that the
results for the two values were similar leads credence to conclusion
that $m$ was taken sufficiently large.

The time the system takes to equilibrate and explore the state space
depends on the inverse temperature parameter $\beta$. The smaller the
$\beta$, the longer it takes.  When $\beta$ is large, it is harder to accept steps, so the
Markov chain will get trapped in a valley where the energy
$\J_{\lambda}$ has a local minimum. Alternatively, when $\beta$ is
small, the Markov chain easily explores the sample space without
settling into a valley. Since we wish to heavily favor the valleys, we
will use a $\beta=0.01$ which we found to be a ``low temperature''
value for the system with most of the probability concentrated at
relatively good minimizers.

To make sure that the Markov chain explores the sample space while
still producing sample districtings of low score, we use a heating
and cooling process. The Markov chain alternates between using a lower
value of $\beta=0.001$ (higher temperature) and the  higher value of $\beta=0.01$ (lower
temperature). If samples are going to be drawn every $m$-th step, then the
value of $\beta$  will switch every $\frac{m}2$ steps, namely
$n=20,000$ when $m=40,000$ or $n=50,000$ when $m=100,000$.
The sample
is taken at the end of each cooling period.

Since the space of redistrictings is enormous, we are only in reality
sampling from a hopefully large region around our initial condition. In
this way all of our samples are related to the current
configuration. The samples produced will be closer  on average to the
current districting than would be redistrictings  chosen randomly according to $\Pr$.
In this way we are tacitly honoring the requirement that the
redistricting have a relation to historical districts if possible.

\begin{figure}[ht]
   \centering\label{fig:calibrating} 
\includegraphics{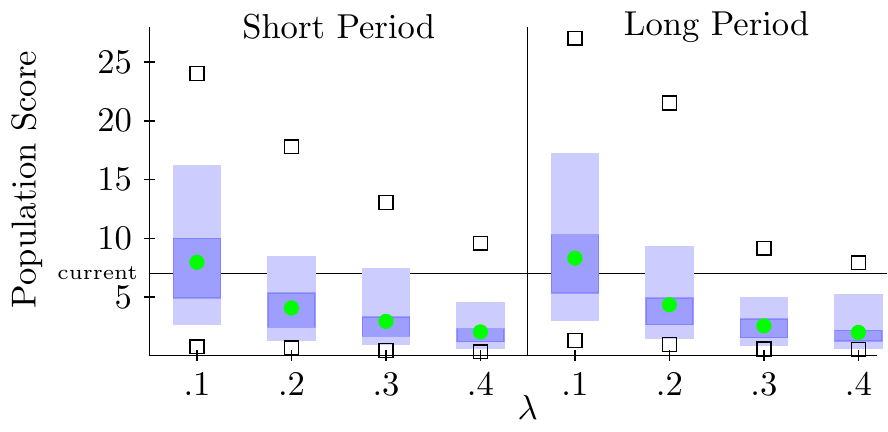}

 \vspace{2em}    

\includegraphics{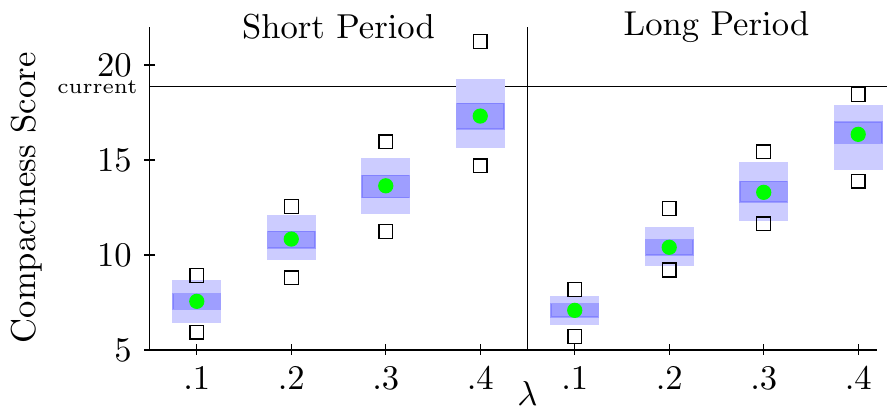}

\caption{Population and Compactness scores  vs 4  values of $\lambda$ each under the shorter and longer heating/cooling cycles.  Solid green dot is the 
  median. Dark blue box is centered on the median to contain 50\%
  of the points. Lighter blue box is centered to contain 90\% of 
  the data. Hollow squares give the max and min values. Solid 
  black horizontal lines give the value of each score for the current redistricting. } 
\end{figure}
\section{Calibrating the parameter $\lambda$}\label{sec:calibrating}
To calibrate the parameter $\lambda$, we tested $\lambda=0.1, 0.2,
0.3$, and $0.4$.  We tune $\lambda$ so that the values of $\JJ_{pop}$
and $\JJ_{compact}$ obtained with $\Pr$ are comparable to those
obtained with the current districts. We also compared the two
different frequencies of cycling between $\beta=0.01$ and
$\beta=0.001$: the ``short period'' of $n=20,000$ and the ``long
period'' of $n=50,000$. The results are given in
Figure~\ref{fig:calibrating}. Since the results are comparable for the
two choices of $n$, we conclude that $n=50,000$ is sufficiently large to
obtain good samples. We would like to choose $\lambda$ so that $\JJ_{pop}$ and $\JJ_{compact}$ are almost always below the current values. This is true for $\lambda=.3$ or $\lambda=.4$. Since our current districting is not very compact but does a good job of evenly dividing the population, we have selected $\lambda=.3$ as our preferred value.

\begin{figure}[ht]\label{fig:racial}
  \centering

\includegraphics{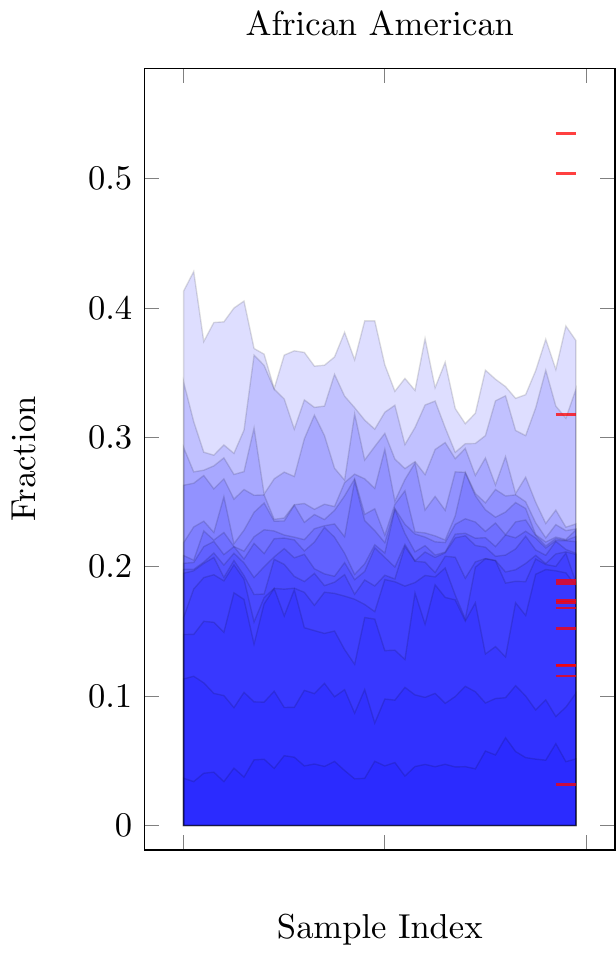}
\includegraphics{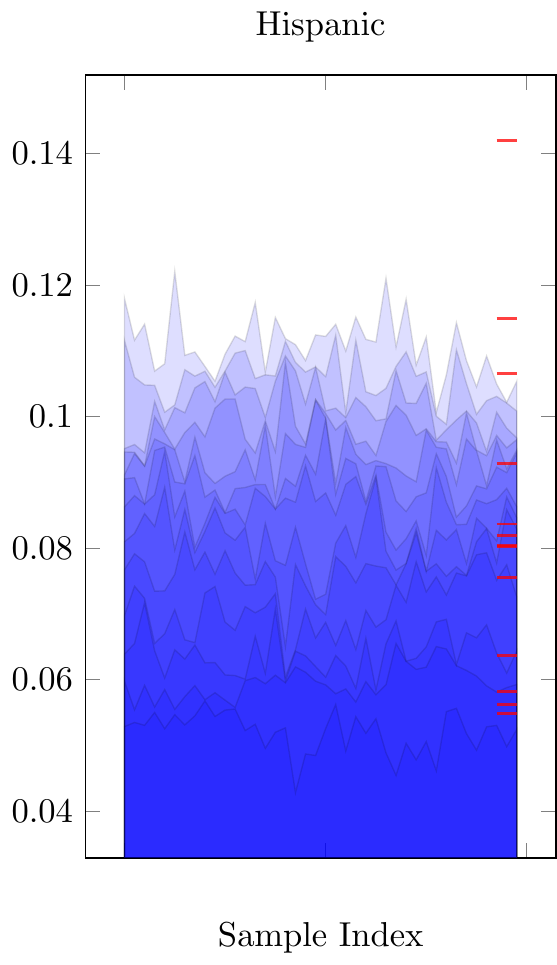}

\caption{Sorted fraction of African Americans and Hispanics in each of
  the districts from the random samples ($\lambda=.3$ and long period). The red
  marks on the left show current redistricting's level.}
\end{figure}

\section{Racial Make Up}\label{sec:racial}
In 2010, about 21\% of the state's population was African American. In
the 2010 and previous redistrictings, an effort was made to produce a
district which could elect African American representatives. Two of
the current districts have 50\% or more African Americans.  Since
African Americans tend to vote Democratic, including any term in the
score function to produce such a district could be perceived as
partisan gerrymandering which we want to avoid in our study.

Nonetheless it is interesting to note the racial makeup of the 
districts created by our model and compare them to the current figures. Figure~\ref{fig:racial} gives the 
results for the African American and Hispanic populations. Our 
samples never produced a majority African American district. The two 
districts with largest African American representation had on average 
around 36\% and 32\% African American population which compares favorably to the 
state wide percentage of 22\%, but not to the current districts. 
The Hispanic population also saw a drop in their percentage in their most
populous district from 14\% in the current districts to on average
11\% in the redistricting produced by our model.

It is of course possible to add additional score functions to prefer
the creation of minority districts. The current results should be seen
as the results if there is no intervention on the policy side to
produce a particular minority district.

\section{Technical Notes}
\label{sec:Technical Notes}
There are a few technical notes that we would like to address in this 
section. In section \ref{sec:ConstructingMeasure}, we define a 
districting as a function $\xi: V \rightarrow \{1,2,\dots,13\}$. To refer to the current congressional districts,
we need each vertex of our graph to belong to only one congressional 
district. We initially defined each vertex as a VTD. However, there 
are $65$ VTDs which lie within two congressional districts. We “split”
each of these VTDs to create two vertices, one for each of the two congressional 
districts. As before, vertices are connected when they are adjacent on 
the map. The congressional district boundaries are used to determine 
the boundary between the split VTDs, allowing us to determine which 
vertices are adjacent. Since population data is only available at the 
VTD level, we approximate the population for a split VTD as half of 
the population of the original VTD. Similarly, we approximate the area 
of a split VTD as half of the area of the original VTD. For brevity,
we refer to each vertex as a VTD, even though some of the vertices are 
a “split” VTD. 

Another technical note is that there are some votes that cannot be
attributed to a specific VTD. For example, absentee voting allows
votes to be cast outside of an individual’s home VTD. The number of
such votes is negligible to those votes that can be attributed to a
VTD and we simply neglect them.

A small error in the MCMC code lead to mildly nonsymmetric transition
probabilities in some of the earlier data collected (in contrast to
what was explained in Section~\ref{Sampling}). Some of that data is
included in the analysis in this preliminary draft. New runs which
will replace all of this data are under way but only partially
complete. Comparison with new data generated shows no qualitative
change from what is shown here. All of the nonsymmetric data will be
replaced in the final version.

\section{Conclusions}\label{sec:conclusions}

We have provided a prototype probability measure on the space of
congressional redistrictings of North Carolina. The measure was
non-partisan in that it considers no information beyond the total
population and shape of the districts. The probability model was then
calibrated to produce redistrictings which are comparable to the
current district to the extent they partition the population equally
and produce compact districts. Then effectively independent draws were
made from this probability distribution using the Metropolis-Hastings
variant of Markov chain Monte Carlo. For each redistricting drawn, the 2012 U.S. House
of representatives election was retabulated using the actual vote
counts to determine the party affiliation of the winner in each
district. The statistics of the number of democratic winners give a
portrait of the range of outcomes possible for the given set of votes
cast. This distribution could be viewed as the true will of the
people.

Redistrictings producing outcomes which are significantly
different than the typical results obtained from random sampled
redistrictings are arguable at odds with the will of the people
expressed in the record of their votes. The fact that the election outcomes are so
dependent on the choice of redistrictings demonstrates the need for
checks and balances to ensure that democracy is served when
redistrictings are drawn and the election outcome is representative of
the votes casted.

It seems unreasonable to expect that politics would not enter into the
process of redistricting. Since the legislators represent the people
and presumably express their will, restricting their ability to
express that will seems contrary to the very idea of democracy. This
seems to be the opinion of a number of the current Supreme Court
Justices. Yet the work in this note could likely be developed into a
criteria to decide when a redistricting fails to be sufficiently
democratic. It would perhaps be reasonable to only allow
redistrictings which yield the more typical results, eschewing the most
atypical as a subversion of the peoples will. This would still leave
plenty of room for politics but add a counter-weight to balance that role of
partisanship when it acts against the democratic ideals of a
republic govened by the people.

\section*{Acknowledgments}
We would like to thank the Duke Math Department and the PRUV
Fellowship program for financial and material support and Austin
M. Graves for assisting in the construction of the graph from the VTD map.

\bibliographystyle{plain}
\bibliography{gerrymandering}

\appendix


\begin{figure}[ht]\centering \label{fig:CurrentDistricting}
\includegraphics[scale=0.52,angle=90]{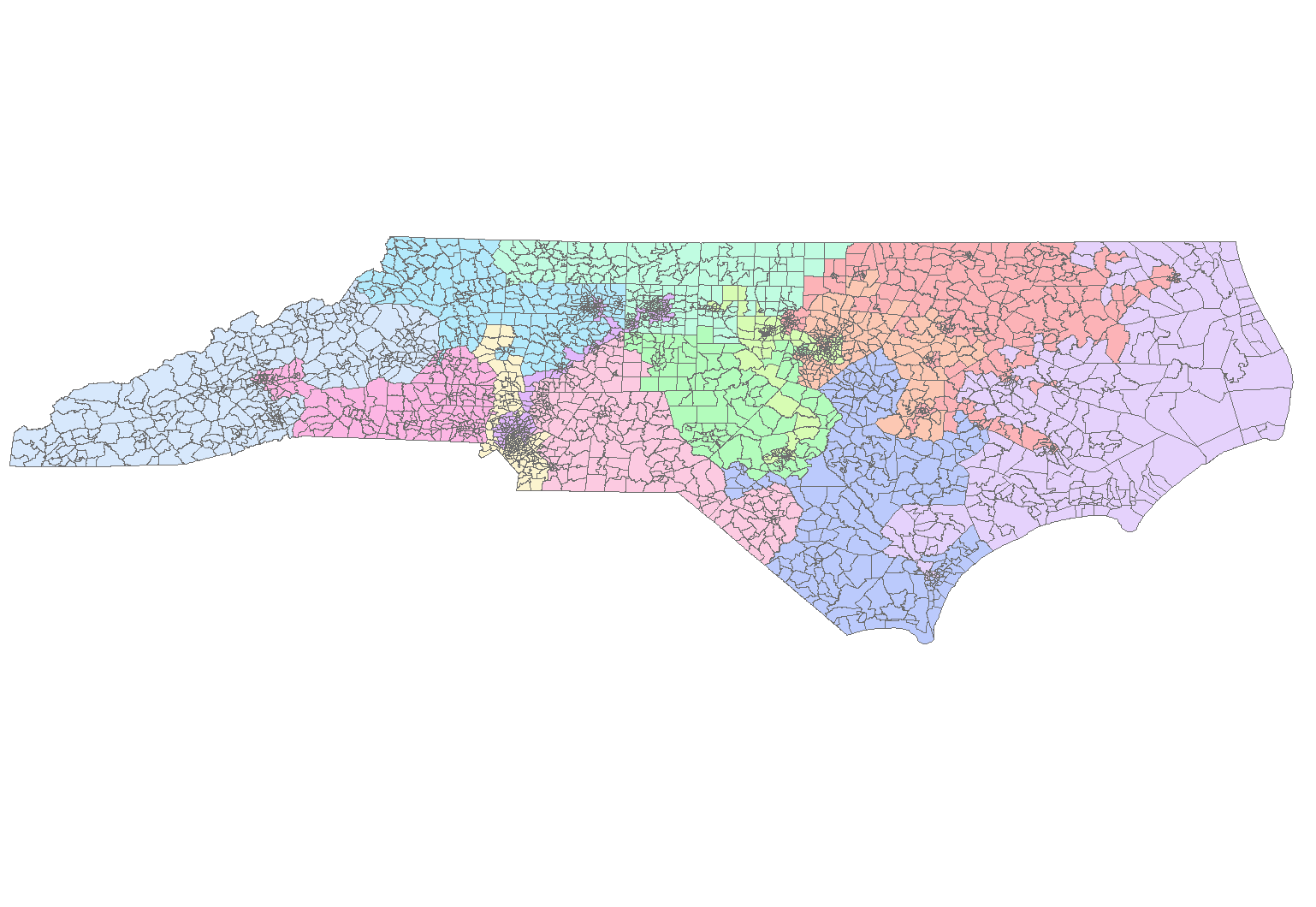} 
\caption{Current districting of NC in 2014}
\end{figure}

\begin{figure}[ht]\centering \label{fig:sampeRedistricting-1}
\includegraphics[scale=0.52,angle=90]{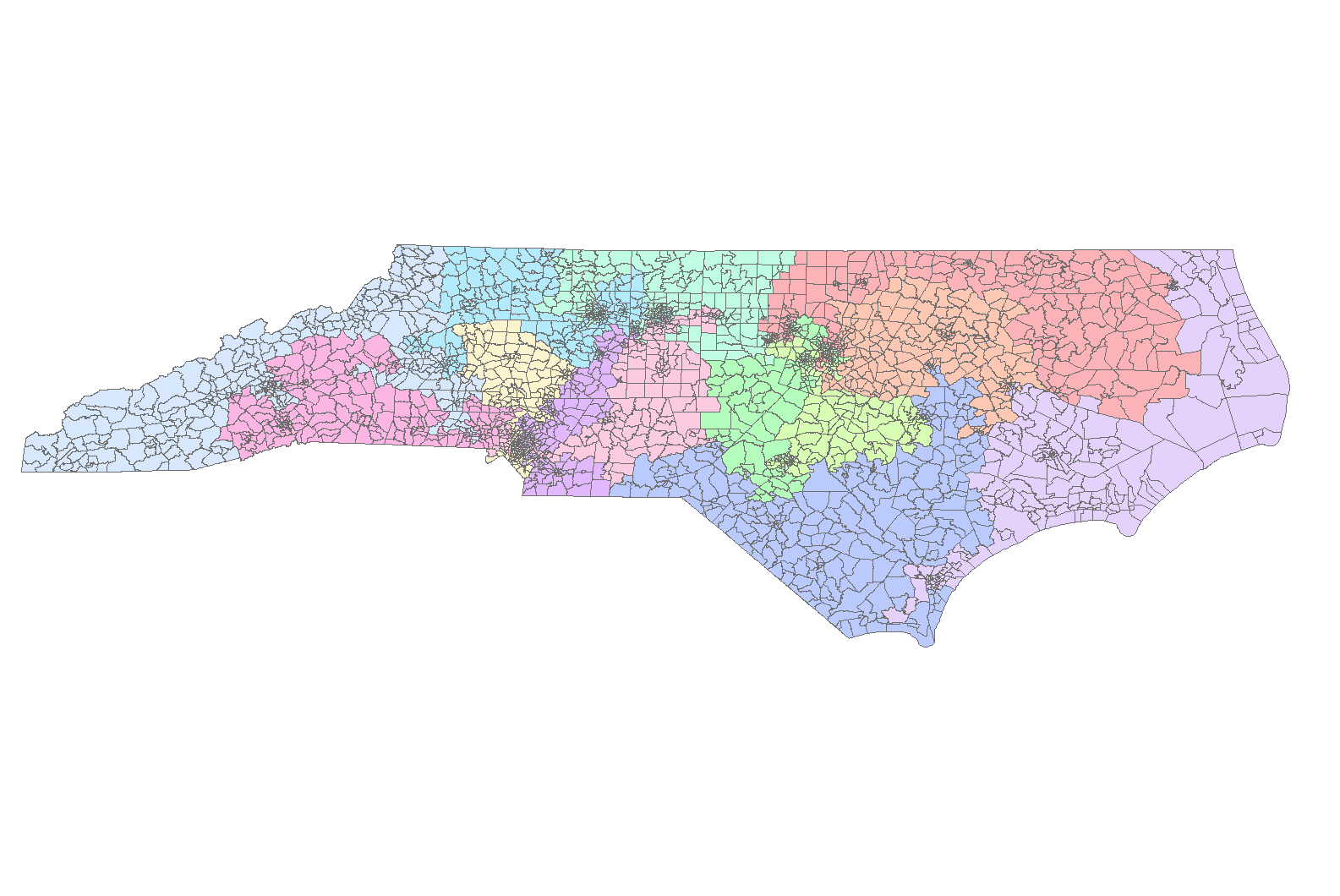} 
\caption{A sample redistricting from $\Pr$ with $\lambda=.3$ and the 
  long heating/cooling cycle.}
\end{figure}

\begin{figure}[ht]\centering \label{fig:sampeRedistricting-2}
\includegraphics[scale=0.52,angle=90]{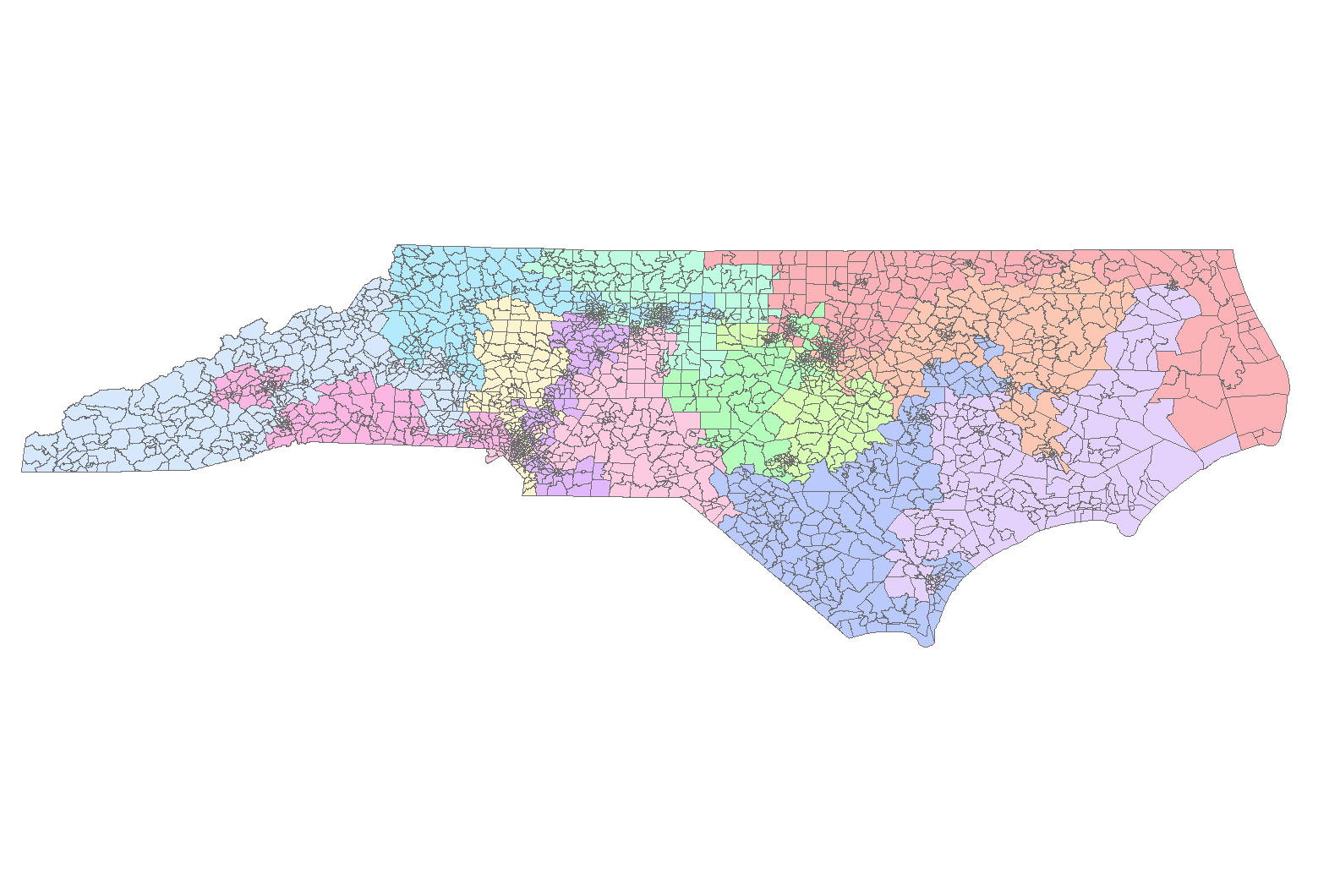}
\caption{A sample redistricting from $\Pr$ with $\lambda=.3$ and the 
  long heating/cooling cycle.}
\end{figure}

\begin{figure}[ht]\centering \label{fig:sampeRedistricting-3}
\includegraphics[scale=0.52,angle=90]{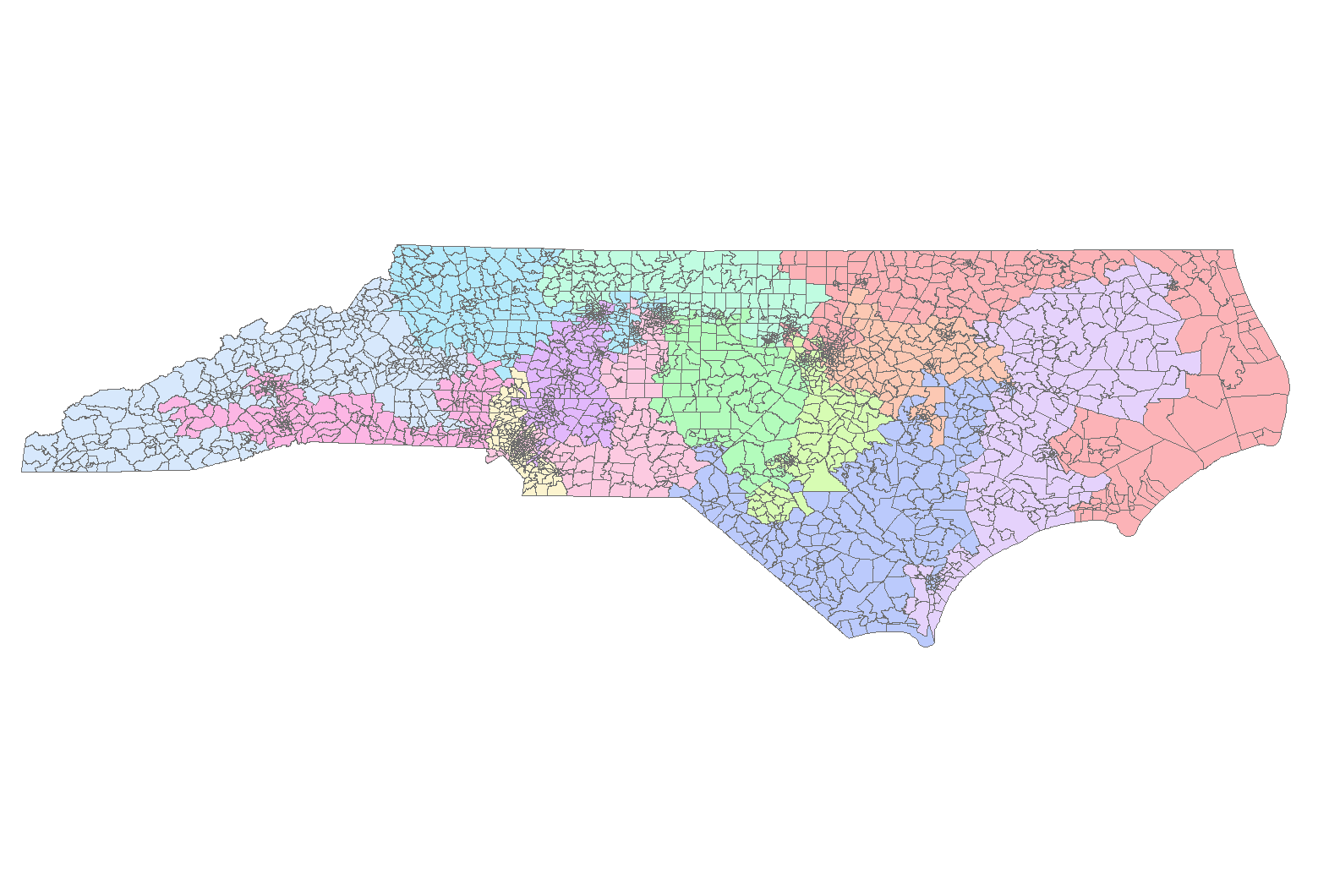} 
\caption{A sample redistricting from $\Pr$ with $\lambda=.3$ and the 
  long heating/cooling cycle.}
\end{figure}

\end{document}